\documentclass[aps,twocolumn,pra,amsmath,amssymb,superscriptaddress,floats,nobibnotes,nofootinbib]{revtex4}

\usepackage{CJK}                     
\usepackage{amsmath}	
\usepackage{gensymb}
\usepackage{hyperref}
\hypersetup{colorlinks=true}
\usepackage{upgreek}
\usepackage{graphics}
\usepackage{hyperref}
\usepackage{epsfig}
\usepackage{color}
\usepackage{bm}
\usepackage{float}
\usepackage{ulem} 
\usepackage{dsfont}                 
\usepackage{wasysym}
\usepackage{multirow}

\usepackage{blindtext}

\usepackage{accents}

\usepackage{graphicx}

\graphicspath{{./}{./plots/}}

\usepackage{url}
\usepackage{multirow}

\newcommand{\upperRomannumeral}[1]{\uppercase\expandafter{\romannumeral#1}}

\begin{document}
\begin{CJK*}{GBK}{song}

\title{Chiral spin state and nematic ferromagnet in the spin-1 Kitaev-$\Gamma$ model} 

\author{Qiang Luo}
\email[]{qiangluo@nuaa.edu.cn}
\affiliation{College of Physics, Nanjing University of Aeronautics and Astronautics, Nanjing, 211106, China}
\affiliation{Key Laboratory of Aerospace Information Materials and Physics (NUAA), MIIT, Nanjing, 211106, China}
\author{Jize Zhao}
\email[]{zhaojz@lzu.edu.cn}
\affiliation{School of Physical Science and Technology $\&$ Key Laboratory of Quantum Theory and Applications of MoE, Lanzhou University, Lanzhou 730000, China}
\affiliation{Lanzhou Center for Theoretical Physics, Key Laboratory of Theoretical Physics of Gansu Province, Lanzhou University, Lanzhou 730000, China}
\author{Jinbin Li}
\affiliation{College of Physics, Nanjing University of Aeronautics and Astronautics, Nanjing, 211106, China}
\affiliation{Key Laboratory of Aerospace Information Materials and Physics (NUAA), MIIT, Nanjing, 211106, China}
\author{Xiaoqun Wang}
\email[]{xiaoqunwang@zju.edu.cn}
\affiliation{School of Physics, Zhejiang University, Hangzhou 310058, China}

\date{\today}

\begin{abstract}
  The higher-spin Kitaev magnets, in which the Kitaev interaction and off-diagonal exchange couplings are overwhelmingly large,
  have emerged as a fertile avenue to explore exotic phases and unusual excitations.
  In this work, we study the quantum phase diagram of the spin-1 Kitaev-$\Gamma$ model on the honeycomb lattice using density-matrix renormalization group.
  It harbours six distinct phases and the intriguing findings are three magnetically ordered phases
  in which both time-reversal symmetry and lattice symmetry albeit of different sort are broken spontaneously.
  The chiral spin state originates from the order-by-disorder effect and exhibits an almost saturated scalar spin chirality at the quantum level.
  Depending on the relative strength of the two interactions, it also features columnar-like or plaquette-like dimer pattern
  as a consequence of the translational symmetry breaking.
  In parallel, the nematic ferromagnets are situated at ferromagnetic Kitaev side and possess small but finite ferromagnetic ordering.
  The lattice-rotational symmetry breaking enforces nonequivalent bond energy along one of the three bonds.
  Although the intrinsic difference between the two nematic ferromagnets remains elusive,
  the discontinuities in the von Neumann entropy, hexagonal plaquette operator, and Wilson loop operator convincingly suggest
  that they are separated via a first-order phase transition.
\end{abstract}

\pacs{}

\maketitle

\section{Introduction}
The celebrated Kitaev honeycomb model \cite{Kitaev2006AP}, which consists of bond-directional Ising couplings,
plays an essential role in the understanding of exotic phases of matter,
e.g., quantum spin liquid (QSL) possessing topological order and fractionalized excitations \cite{Wen2019NPJ,Rousochatzakis2024RoPP}.
The QSL is a nonmagnetic phase which escapes from any spontaneous symmetry breaking down to the lowest temperature \cite{Balents2010Nature}
and has an intimate relation to high-temperature superconductivity in cuprates and quantum computations.
In this endeavor, great efforts have been made to realize the appealing Kitaev interaction in real materials.
Among them, Jackeli and Khaliulin pointed out that the strong spin-orbit coupling together with proper electronic correlations
could give rise to this interaction \cite{Jackeli2009PRL}.
In the quest for these ``Kitaev materials" on the honeycomb lattice,
which include iridates \cite{Ye2012PRB,Choi2012PRL,Chun2015NP},
$\alpha$-RuCl$_3$ \cite{Sears2015PRB,Sears2017PRB,Baek2017PRL,Wolter2017PRB,Wang2017PRL,Zheng2017PRL,Do2017NP,Ran2017PRL,WinterNcom2018},
and Na$_2$Co$_2$TeO$_6$ and Na$_3$Co$_2$SbO$_6$ \cite{Lin2021NC,Li2022PRX,Takeda2022PRR,Guang2023PRB},
it turns out that they usually display magnetic orderings at low temperatures for the possible existence of competing interactions.
In spite of this, signatures of continuum spectrum \cite{Banerjee2016NM} and half-quantized thermal Hall conductivity \cite{Kasahara2008Nature} have been reported in $\alpha$-RuCl$_3$,
while an intriguing triple-$Q$ order that is proximity to hidden SU(2) point is studied in Na$_2$Co$_2$TeO$_6$ \cite{Kruger2023PRL}.

On the other hand, devising alternative Kitaev materials with $S > 1/2$ could potentially expand the field of seeking interesting phases.
So far, the highly desirable Kitaev interaction has been identified in some higher-spin magnets on the honeycomb lattice \cite{Stavropoulos2019PRL}.
In the spin-1 Ni-based antimonates $A_3$Ni$_2$SbO$_6$ ($A$ = Li, Na), the thermal entropy of $\sim\frac12\ln3$ per Ni element over the Neel temperatures
firmly hints the existence of the Kitaev interaction \cite{Zvereva2015PRB}.
In their allotropic Na$_3$Ni$_2$BiO$_6$, the Kitaev interaction is proposed to enhance exchange frustration
and stabilize the one-third magnetization plateau in the partial spin-flop ferrimagnetic order \cite{Shangguan2023NP}.
In KNiAsO$_4$, inclusion of the Kitaev interaction is essential to adequately explain the zigzag magnetic order and other thermodynamical quantities \cite{Taddei2023PRR}.
In the spin-$3/2$ Cr-based van der Waals magnet CrI$_3$, CrSiTe$_3$, and CrGeTe$_3$ \cite{Xu2018npjCM,Xu2020PRL,Stav2021PRR},
the Kitaev interaction is found to be induced by heavy ligands of I/Te despite a weak spin-orbit coupling related to Cr irons.
Although the compelling evidences to advocate the Kitaev QSL remain unclear \cite{Zhou2021PRB},
there are proposals to impair the non-Kitaev terms in CrSiTe$_3$ and CrGeTe$_3$ by strain engineering \cite{Xu2020PRL}.
In addition, identification of the Kitaev interaction in triangular-lattice compounds has also been reported,
which include the spin-1 NiI$_2$ \cite{Song2022Nature} and the spin-$3/2$ $1T$-CrTe$_2$ \cite{Freitas2015JPCM}.
In the latter, the Kitaev interaction is crucial to understand the orientation of the magnetic moment
that is 70$^{\circ}$ off the normal direction of the plane \cite{Huang2023PRB}.

According to the proposal by Baskaran, Sen, and Shankar \cite{Baskaran2008PRB},
the extensive number of local conserved quantities guarantee the vanishing of the spin-spin correlation functions beyond the nearest-neighbor bonds
in the pure Kitaev limit.
Therefore, the ground state of the spin-$S$ Kitaev model should always be disordered, with the possibility of being a sort of QSL.
Hitherto, the specific heat and thermal entropy \cite{Koga2018JPSJ,Oitmaa2018PRB}, the topological entanglement entropy \cite{Lee2020PRR},
and the excitation spectrum \cite{Chen2022PRB} of the spin-1 Kitaev QSL have been widely studied.
A magnetic-field-induced intermediate nonmagnetic phase sandwiched between the Kitaev QSL and the polarized phase is reported \cite{Zhu2020PRR,Hickey2020PRR,Khait2021PRR}.
In the spin-$3/2$ analogy, the spin can be reexpressed by the SO(6) Majorana representation and the Kitaev QSL is found to couple to a static $\mathbb{Z}_2$ gauge field \cite{Jin2022NC}.
Recently, there has been a surge interest in higher-spin models with Kitaev-type interaction.
In the spin-1 Kitaev model, it is found that the Kitaev QSL survives against finite Heisenberg interaction \cite{Dong2020PRB},
and intervals of the Kitaev QSLs shrink rapidly as $S$ increases \cite{Fukui2022PRB}.
Furthermore, the interplay of Kitaev interaction and other competing terms,
e.g., biquadratic exchange \cite{Pohle2023PRB} and single-ion anisotropy \cite{Singhania2023arXiv},
is anticipated to promote inspiring phases and unconventional quantum phase transitions (QPTs).

Motivated by the Kitaev materials with $j_{\rm eff} = 1/2$, the Kitaev-$\Gamma$ model is broadly recognized as their minimal model \cite{Wang2017PRB}.
Despite great efforts, the full quantum phase diagram of this model remains unsettled, especially when $K$ and $\Gamma$ have opposite signs \cite{Rousochatzakis2024RoPP,Rau2014PRL,Wang2019PRL,Luo2021NPJ,Lee2020NC,Gohlke2020PRR}.
As $S > 1/2$, the quantum fluctuations can be suppressed in some sense, leaving the possibility of figuring out a more precise quantum phase diagram.
Although relevant real materials are not available,
our work serves as a reference to understand the novel phases in other higher-spin Kitaev magnets.
By virtue of the density-matrix renormalization group (DMRG) method \cite{White1992PRL,Peschel1999,Schollwock2005RMP},
we elucidate the quantum phase diagram by systematically changing the ratio between the Kitaev and $\Gamma$ interactions.
We uncover six phases in total, in which two of them are ferromagnetic (FM) and antiferromagnetic (AFM) Kitaev QSLs and
one is a possible nonmagnetic AFM $\Gamma$ (A$\Gamma$) phase that is surrounded in the $\Gamma$ limit.
In addition, there are three magnetically ordered states, which are the chiral spin state, and two emergent nematic ferromagnets near the FM Kitaev QSL.

The layout of the rest of the paper is as follows.
In Sec.~\ref{SEC:Model}, we explain the model and the associated operators, introduce the numerical methods, and present the quantum phase diagram.
The detailed analysis of the chiral spin state and the nematic ferromagnets are shown in Sec.~\ref{SEC:Chiral} and Sec.~\ref{SEC:NmtcFM}, respectively.
Finally, conclusion and some further discussion are given in Sec.~\ref{SEC:CONC}.

\begin{figure}[!ht]
  \centering
  \includegraphics[width=0.95\columnwidth, clip]{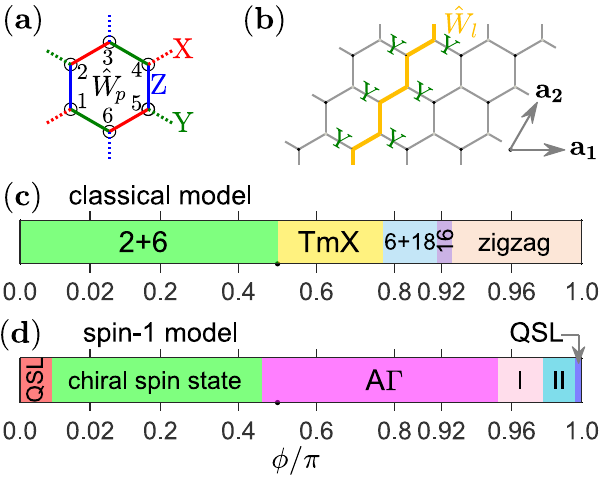}\\
  \caption{(a) Illustration of the hexagonal plaquette operator $\hat{W}_p$. The nearest-neighbor links are distinguished as $\textbf{X}$ (red), $\textbf{Y}$ (green), and $\textbf{Z}$ (blue) bonds, respectively.
  (b) Illustration of the Wilson loop operator $\hat{W}_l$. The yellow loop denotes the Wilson loop operator along the $\textbf{a}_2 (1/2, \sqrt{3}/2)$ direction.
  (c) The classical phase diagram of the Kitaev-$\Gamma$ model, based on previous Monte Carlo results \cite{Rayyan2021PRB,Chen2023NJP}.
  Here, the integer indicate the number of sublattices within a unit cell of the corresponding magnetic order,
  and the TmX denotes the 18-site triple-meron crystal.
  (d) The quantum phase diagram of the spin-1 Kitaev-$\Gamma$ model.
  The acronym A$\Gamma$ stands for an antiferromagnetic $\Gamma$ phase,
  and the symbols I and II represent two nematic ferromagnets (see the main text for details).
  }\label{FIG-Hex24}
\end{figure}

\begin{figure*}[htb]
  \centering
  \includegraphics[width=2.0\columnwidth, clip]{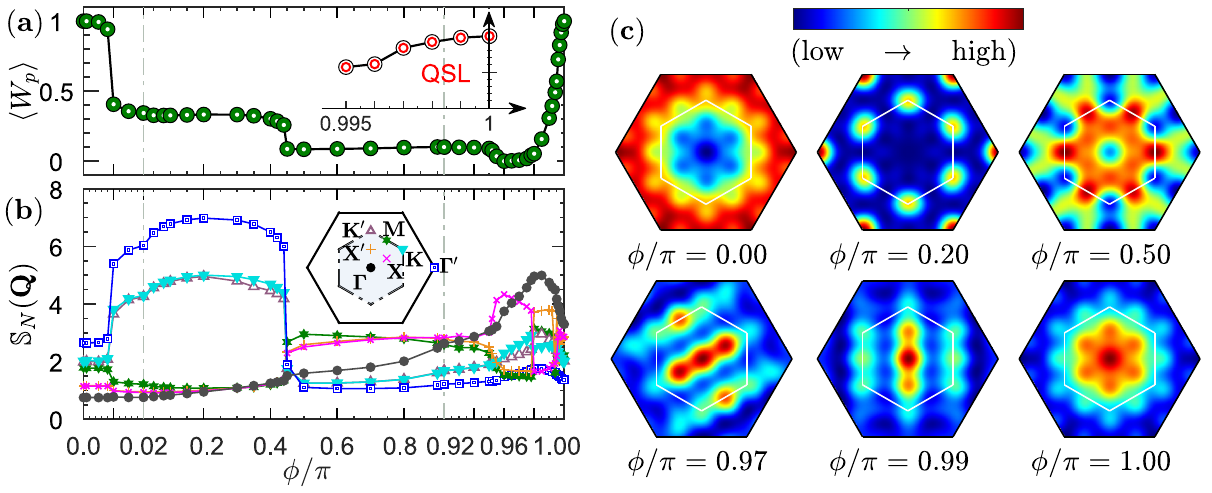}\\
  \caption{(a) The flux-like density $\langle\overline{W}_p\rangle$ as a function of $\phi$ on the 24-site hexagonal cluster.
    Inset: Zoom-in of $\langle\overline{W}_p\rangle$ near the FM Kitaev limit.
    (b) The SSF $\mathbb{S}_N({\bf{q}})$ at high-symmetry points $\boldsymbol{\Gamma}$ (black), in the reciprocal space (see inset).
    The vertical dash-dotted gray lines represent the dividing lines separating three regions that have different plotting scales,
    which are ${\phi}/{\pi} \in [0,0.02]$, ${\phi}/{\pi} \in [0.02,0.92]$, and ${\phi}/{\pi} \in [0.92, 1]$.
    (c) Landscapes of representative SSF at $\phi/\pi$ = 0.00, 0.20, 0.50, 0.97, 0.99, and 1.00, respectively.
    The color scale in each panel is normalized individually to its own max intensity.
    The asymmetric SSFs at $\phi/\pi$ = 0.97 and 0.99 may relate the broken $C_3$ symmetry in the nematic ferromagnets.
    }\label{FIG-WpSSF}
\end{figure*}

\section{Model and Phase Diagram}\label{SEC:Model}
We study the spin-1 Kitaev-$\Gamma$ model on the honeycomb lattice, which has the Hamiltonian defined as
\begin{align}\label{EQ:KGHam}
\mathcal{H} =
    & \sum_{\left<ij\right>\parallel\gamma} \Big[K S_i^{\gamma} S_j^{\gamma}
    + \Gamma \big(S_i^{\alpha}S_j^{\beta}+S_i^{\beta}S_j^{\alpha}\big)\Big],
\end{align}
where $S_i^{\gamma}$~($\gamma$ = $x$, $y$, and $z$) is the $\gamma$-component of a spin operator at site $i$
and $\left<ij\right>\parallel\gamma$ denotes the $\gamma$-type nearest-neighbor bond connecting sites $i$ and $j$.
$\alpha$ and $\beta$ are the two remaining bonds under a cyclic permutation of $\{x, y, z\}$.
In addition, $K$ and $\Gamma$ stand for the Kitaev and $\Gamma$ interactions, respectively.
We consider a trigonometric parametrization of these interactions such that
$K = \cos\phi$ and $\Gamma = \sin\phi$ with $0 \leq \phi \leq \pi$.
We note that this parameter region has been widely studied in its spin-1/2 analogy
where $\Gamma$ term is widely found to be positive in existing Kitaev materials,
and our study is beneficial for mapping out the global phase diagram in extended Kitaev models in the future.

We have studied the quantum phase diagram of this model by using the DMRG method on two different clusters.
One is a $C_3$-symmetric 24-site hexagonal cluster,
which is primarily adopted since it retains the symmetry of the Hamiltonian in Eq.~\eqref{EQ:KGHam}.
The other is a $2 \times L_x \times L_y$ rhombic cluster with $N = 2L_xL_y$ sites in total,
which is utilized for comparison and extrapolation.
Full periodic boundary conditions are used in both clusters, with the exception of the long cylinder in the latter case.
During the computation, we keep as many as $m = 4000$ block states and exert up to 24 sweeps until the typical truncation error is less than $10^{-6}$.

Different from the spin-$1/2$ Kitaev model, the spin-1 counterpart does not seem to have an exact solution up to now.
Nevertheless, it has been verified that this model can be characterized by a couple of conserved $\mathbb{Z}_2$ quantities.
One of such quantities is the hexagonal plaquette operator \cite{Baskaran2008PRB}
\begin{equation}\label{EQ:PlqttWp}
\hat{W}_p = e^{\imath\pi(S_1^x + S_2^y + S_3^z + S_4^x + S_5^y + S_6^z)},
\end{equation}
where $S_j^{\gamma}$ is the outgoing bond around the hexagon path at site $j$, see Fig.~\ref{FIG-Hex24}(a).
Quite recently, by introducing a parton construction with $8S$ Majorana fermions,
Ma shows that the $\hat{W}_p$s are still $\mathbb{Z}_2$ gauge fluxes \cite{Ma2023PRL}.
Another representative quantity is the Wilson loop operator
\begin{equation}\label{EQ:WilsonLoopWl}
\hat{W}_l = \prod_{j=1}^{2L_y} e^{\imath\pi S_j^y},
\end{equation}
where $S_j^y$ is the outgoing bond along the circle path at site $j$, see Fig.~\ref{FIG-Hex24}(b).
It is straightforward to confirm that both operators commute with the Hamiltonian of the Kitaev model
and their eigenvalues in the ground state are numerically found to be 1.
While the nature of the excitations remains elusive
\cite{Koga2018JPSJ,Oitmaa2018PRB,Lee2020PRR,Chen2022PRB,Zhu2020PRR,Hickey2020PRR,Khait2021PRR},
it is widely believed that the ground state of the spin-1 Kitaev model is a sort of QSL
in which the spin-spin correlation functions $\langle\textbf{S}_i\cdot\textbf{S}_j\rangle$ disappear beyond the nearest-neighbor bonds.
The QSL can also be ascertained by the static structure factor (SSF),
which features a diffusive pattern in the reciprocal space.
The SSF is given by $\mathbb{S}_N({\bf{q}}) = \sum_{\alpha\beta}\delta_{\alpha\beta}\mathbb{S}_N^{\alpha\beta}({\bf{q}})$, where
\begin{equation}\label{EQ:SSF}
\mathbb{S}_N^{\alpha\beta}({\bf{q}})=\frac{1}{N}\sum_{ij} \langle{S^{\alpha}_i {S^{\beta}_j}}\rangle e^{\imath{\bf{q}}\cdot{({\bm{R}}_i-{\bm{R}}_j)}}.
\end{equation}
For the magnetically ordered state, the magnetic order parameter is defined as $M_N({\bf{Q}}) = \sqrt{\mathbb{S}_N({\bf{Q}})/N}$
with ${\bf{Q}}$ being the ordering wavevector.

\begin{figure*}[htb]
\centering
  \includegraphics[width=2.0\columnwidth, clip]{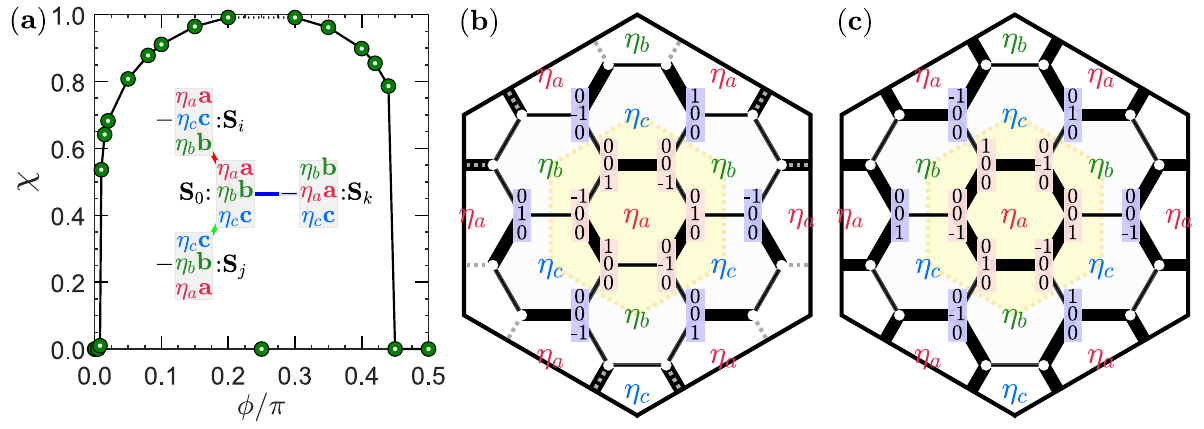}\\
  \caption{(a) The scalar spin chirality $\chi$ as a function of $\phi$ on the 24-site hexagonal cluster. Inset: Illustration of the spin patterns on the center spin $\mathbf{S}_0$ and its three nearest-neighbor spins $\{\mathbf{S}_i, \mathbf{S}_j, \mathbf{S}_k\}$.
  (b) and (c) sketch the dimer patterns of the columnar-like and plaquette-like chiral spin state at $\phi/\pi = 0.2$ and $\phi/\pi = 0.4$, respectively.
  The wider bond has a higher intensity of the bond energy.}\label{FIG-ChiralVBS}
\end{figure*}

With the inclusion of the $\Gamma$ term, the ground state is more involved and the DMRG method is amenable to figure out the quantum phase diagram.
Before we embark on presenting the details, it is worthwhile to first recall the classical phase diagram of the Kitaev-$\Gamma$ model, see Fig.~\ref{FIG-Hex24}(c).
According to the previous Monte Carlo simulations, there are five large-unit-cell orderings \cite{Rayyan2021PRB,Chen2023NJP}.
These incorporate the (2+6)-site order when $0 < \phi/\pi < 1/2$,
and the triple-meron crystal with eighteen sites, the (6+18)-site order, the 16-site order, and the four-site zigzag order when $1/2 < \phi/\pi < 1$.
In addition, a recent study also discoveries an incommensurate phase that is sandwiched between the triple-meron crystal and (6+18)-site order \cite{Stavropoulos2023arXiv}.

As can be seen from Fig.~\ref{FIG-WpSSF}(a), the Kitaev QSLs in both FM and AFM Kitaev limits are extremely vulnerable and they are ruined by a tiny $\Gamma$ interaction.
Guided by the flux-like density $\langle\overline{W}_p\rangle$ and other quantities,
we conclude that the QPTs from Kitaev QSLs to their neighboring phases occur when $\phi/\pi \approx$ 0.009 and 0.996, respectively.
This indicates that the interval of the FM Kitaev QSL is smaller than its AFM analogy.
We note that such an asymmetric stability of the Kitaev QSLs may relate to the interplay of the two flux-pair processes
in which their magnitudes rely on the sign of the Kitaev interaction \cite{Zhang2021PRB}.
By adding the $\Gamma$ interaction in the AFM Kitaev side, the ground state turns to be a magnetically ordered chiral spin state
which occupies a significant portion of the phase diagram at $\phi/\pi \in (0.009, 0.445)$.
It has a $1/3$ flux density and possesses pronounced peaks at $\textbf{K}$ and $\boldsymbol{\Gamma}'$ points
(corners of the first and second Brillouin zones, respectively) in the reciprocal space.
By contrast, when considering the $\Gamma$ interaction in the FM Kitaev side,
there are two emergent nematic ferromagnets which break lattice-rotational symmetry and time-reversal symmetry simultaneously
but preserve translational invariance.
The $C_3$ rotational symmetry breaking is manifested by the discrepant bond energies along the three bond directions.
In addition, the peak at the $\boldsymbol{\Gamma}$ point is sizable and it survives even though the system size is infinite.
The discrepancy of the two may lie in the orientation of the nematicity, which causes to different subleading peaks in the Brillouin zone, see Fig.~\ref{FIG-WpSSF}(c).
The remaining phase in the phase diagram is called A$\Gamma$ phase.
We stress that it is merely one phase rather than a complicated region containing a few ones.
Based on the behavior of the SSF at different system sizes and its comparison to the spin-$1/2$ analogy (see Appendix~\ref{appendixA}),
we propose that it is a plausible disordered phase.
However, the nature of A$\Gamma$ phase is beyond the scope of this work and deserves deeper investigation by large-scale many-body computations.

\section{Chiral spin state}\label{SEC:Chiral}
\subsection{Order-by-disorder effect}
We start by considering the unfrustrated case where $K, \Gamma > 0$ at the classical level.
First of all, it is helpful to recall the hidden SU(2) Heisenberg point at $K = \Gamma$ \cite{Chaloupka2015PRB},
which provides a natural starting point for understanding the magnetic ordering.
To disclose the SU(2) point, one needs to employ a six-sublatice $\mathcal{T}_6$ transformation,
which rotates the six spins in a unit cell according to the protocol
$(\tilde{S}_1^x, \tilde{S}_1^y, \tilde{S}_1^z) = (S_1^x, S_1^y, S_1^z)$,
$(\tilde{S}_2^x, \tilde{S}_2^y, \tilde{S}_2^z) = (-S_2^y, -S_2^x, -S_2^z)$,
$(\tilde{S}_3^x, \tilde{S}_3^y, \tilde{S}_3^z) = (S_3^y, S_3^z, S_3^x)$,
$(\tilde{S}_4^x, \tilde{S}_4^y, \tilde{S}_4^z) = (-S_4^x, -S_4^z, -S_4^y)$,
$(\tilde{S}_5^x, \tilde{S}_5^y, \tilde{S}_5^z) = (S_5^z, S_5^x, S_5^y)$, and
$(\tilde{S}_6^x, \tilde{S}_6^y, \tilde{S}_6^z) = (-S_6^z, -S_6^y, -S_6^x)$.
Thus, the essential of this transformation is cyclic (anticyclic) permutations among the spin components at odd (even) sites.
Therefore, while the ground state turns out to be a dual FM phase after the rotation,
it is dubbed counter-rotating spiral order in the original basis \cite{Stavropoulos2018PRB},
in which the spins on one of the two sublattices rotate in the clockwise direction
while the remaining spins on the other sublattice rotate counter-clockwise around the [111] axis.

According to the $\mathcal{T}_6$ transformation, the Hamiltonian in Eq.~\eqref{EQ:KGHam} can be rewritten as
\begin{align}\label{EQ:KGHamT6}
\tilde{\mathcal{H}} = & -\Gamma \sum_{\left<ij\right>} \mathrm{\tilde{S}}_i\cdot \mathrm{\tilde{S}}_j
                + (\Gamma-K)\sum_{\left<ij\right>\parallel\gamma} \tilde{S}_i^{\gamma} \tilde{S}_j^{\gamma}
\end{align}
where $\gamma$ stands for the type of Ising bond.
The first term in Eq.~\eqref{EQ:KGHamT6} is the FM Heisenberg interaction, while the second is the Kitaev interaction.
However, an interesting observation of the $\mathcal{T}_6$ transformation is that it can redistribute types of Ising bonds,
reformulating a Kekul\'{e}-type pattern.
Specifically, the unit cell shown in Fig.~\ref{FIG-Hex24}(a) is recast into
\begin{equation}\label{EQ:HexKeKule}
\includegraphics[width=0.66\columnwidth, clip]{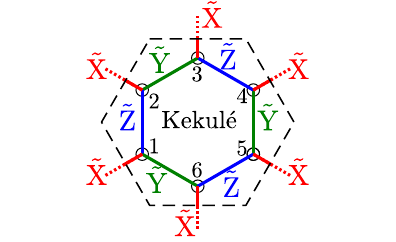}
\end{equation}
Here, $\mathbf{\tilde{X}}$ bonds orient outside, and $\mathbf{\tilde{Y}}$ and $\mathbf{\tilde{Z}}$ bonds alternate along the benzene-like ring.
Supposed that one of the spin components is the largest (which will be justified later), saying $\tilde{S}^x$,
then the bond energy along $\mathbf{\tilde{X}}$ bonds should be either the largest or the smallest, depending on the sign of $(\Gamma-K)$.
When $\Gamma$ is larger, it is detrimental to the bond energy along $\mathbf{\tilde{X}}$ bonds, resulting in a plaquette-like bond energy pattern.
By contrast, a columnar-like bond energy pattern should be achieved when $K$ is larger.

In what follows, we will advocate the plaquette-like and columnar-like bond energy patterns by considering quantum fluctuations over the classical configurations.
Following the recipe proposed by Rousochatzakis \textit{et al.} \cite{Rousochatzakis2017PRL,Rousochatzakis2018NC}, we can choose and parametrize a central spin as $\mathbf{S}_0 = (\eta_a a, \eta_b b, \eta_c c)$,
where ($\eta_a$, $a$) \big[respectively, ($\eta_b$, $b$) and ($\eta_c$, $c$)\big] stand for the sign and intensity of $S_0^x$ (respectively, $S_0^y$ and $S_0^z$).
Here, $a^2 + b^2 + c^2 = S^2$ and $\eta$s are independent Ising variables that can be $\pm1$.
By noticing the peculiar structures of the Kitaev and $\Gamma$ interactions,
the three neighbor spins subject to different bond types can be determined as
$\mathbf{S}_i = -(\eta_a a, \eta_c c, \eta_b b)$, $\mathbf{S}_j = -(\eta_c c, \eta_b b, \eta_a a)$, and $\mathbf{S}_k = -(\eta_b b, \eta_a a, \eta_c c)$
[see the inset of Fig.~\ref{FIG-ChiralVBS}(a)].
In doing so, all the spins can be determined gradually, formulating specific spin textures with finite unit cells.
Given that there are only three different $\eta$s, the ground-state degeneracy is eight in addition to the free choice of $\{a, b, c\}$.
Typically, the unit cell contains six spins except for the case of two-sublattice AFM phase in which $\eta_a = \eta_b = \eta_c = \pm1$ and $a = b = c = S/\sqrt3$.
It is in this sense that the phase is termed (2+6)-ordering classically.

Irrelevant of the choice of $\{\eta_a, \eta_b, \eta_c\}$ and $\{a, b, c\}$, the ground-state energy per site is $e_{cl} = -(\Gamma + K/2)S^2$ classically.
When the fluctuations are involved, the real-space perturbation theory is amenable to address the issue of choosing $\{a, b, c\}$.
The second-order calculation shows that the leading energy correlation is \cite{Rousochatzakis2024RoPP,Rousochatzakis2020KITP}
\begin{equation}\label{EQ:RSPTEg}
e_{cl}^{(2)} = -\frac{(\Gamma-K)^2S}{32|\Gamma+2K|}\left[(a/S)^4 + (b/S)^4 + (c/S)^4\right].
\end{equation}
Thus, when $K \neq \Gamma$, the total energy is minimized as long as one of the elements in $\{a, b, c\}$ is unitary while the rest are zero.
Meanwhile, only one of the $\eta$s that adheres to the nonzero element survives.
This leads to the so-called Cartesian states which map to dimer coverings of the honeycomb lattice \cite{Baskaran2008PRB}.

\begin{figure}[!ht]
\centering
  \includegraphics[width=0.95\columnwidth, clip]{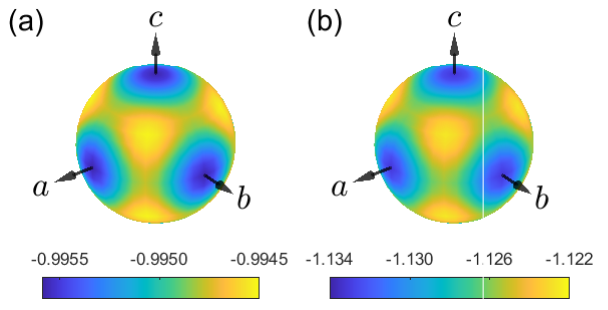}\\
  \caption{Spherical plots of the spin-wave energy $\epsilon_{sw}$ at (a) $\phi/\pi = 0.2$ and (b) $\phi/\pi = 0.4$, respectively.
  The central spin $\mathbf{S}_0$ is parameterized as $\left(S^x, S^y, S^z\right) = \left(\sin\vartheta\cos\varphi, \sin\vartheta\sin\varphi, \cos\vartheta\right)$.
  }\label{FIG-SpinWaveEg}
\end{figure}

The order-by-disorder effect can also be verified by the linear spin-wave theory.
Using the Holstein-Primakoff transformation,
the spin operator at $i$-th site can be approximately rewritten as
$\hat{\textbf{S}}_i = \left(\sqrt{S/2}(b_i+b_i^{\dagger}), -\imath\sqrt{S/2}(b_i-b_i^{\dagger}), S-b_i^{\dagger}b_i\right)$,
and the Hamiltonian in Eq.~\eqref{EQ:KGHam} is readily to be expressed into a quadratic form.
The spin-wave energy is given by \cite{Luo2021NPJ}
\begin{align}\label{EQ:SWEg}
e_{sw} = \left(1+\frac{1}{S}\right)e_{cl} + \frac{S}{2}\sum_{\upsilon\bf{q}} \omega_{\upsilon\bf{q}}.
\end{align}
When the exchange couplings are fixed, the spin-wave energy $e_{sw}$ can vary slightly for different choice of $\{a, b, c\}$.
Figure~\ref{FIG-SpinWaveEg} shows the spherical plots of the spin-wave energy for different values of $\{a, b, c\}$ at $\phi/\pi$ = 0.2 and 0.4.
In both cases, the energy takes the minimal value on the points where one of the three spin components is $\pm S$.
This result is fairly consistent with that of the real-space perturbation theory shown in Eq.~\eqref{EQ:RSPTEg}.
As a byproduct, spin-wave energy gains a maximum when $|a| = |b| = |c| = S/\sqrt{3}$,
showing that the AFM-like phase is not favored by quantum fluctuations.

\begin{figure*}[htb]
\centering
  \includegraphics[width=0.90\linewidth, clip]{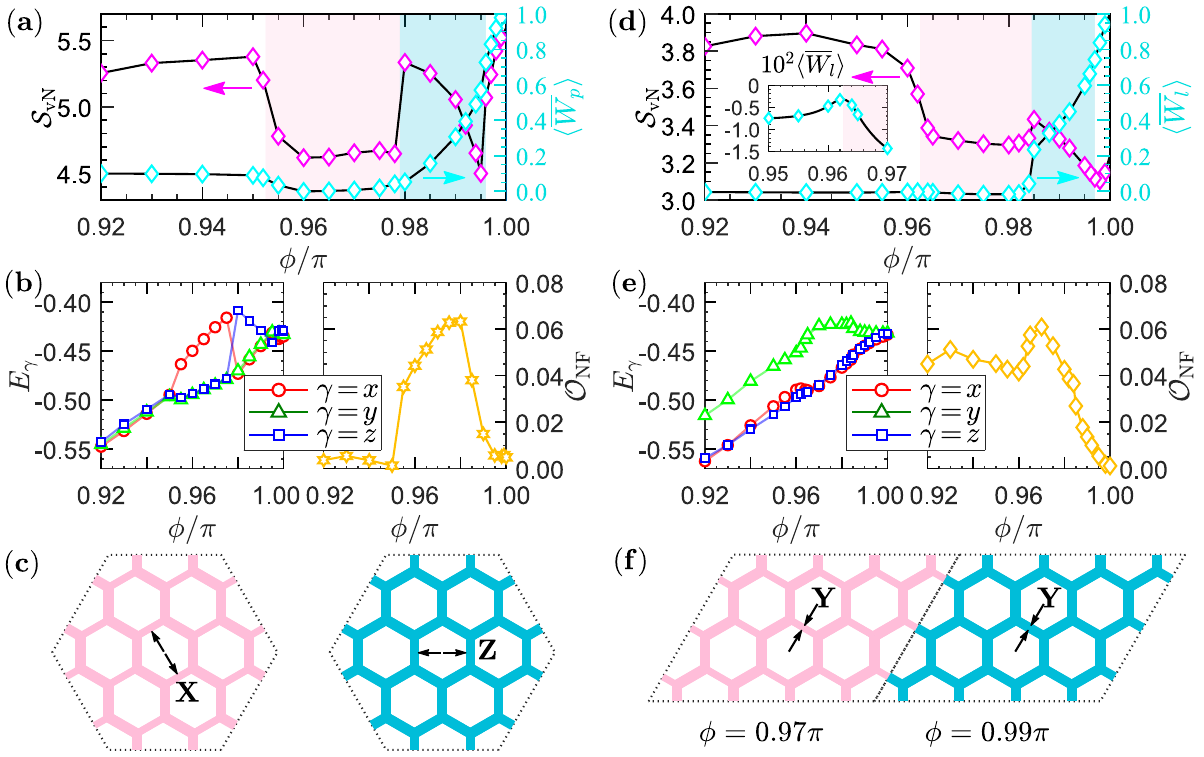}\\
  \caption{(a) The von Neumann entropy $\mathcal{S}_{\rm vN}$ (left axis) and the flux-like density $\langle\overline{W}_p\rangle$ (right axis) as a function of $\phi/\pi \in [0.92, 1.00]$ on the 24-site hexagonal cluster. The shaded areas denote the regions of the two nematic ferromagnets.
  (b) The bond energy $E_{\gamma}$ [$\gamma$ = $x$ (red), $y$ (green), and $z$ (blue)] (left panel) and the nematic order parameter $\mathcal{O}_{\rm NF}$ (right panel) in the regime of interest.
  (c) Illustration of the patterns of the nematic ferromagnets for $\phi/\pi$ = 0.97 (left panel) and $\phi/\pi$ = 0.99.
  The thickness of the bond is proportional to cube of the relativity bond energy $E_{ij}$, i.e., $\propto (E_{ij}/\max(E_{ij}))^3$, for the guide of eyes.
  (d)-(f) and the same as these of (a)-(c) but for the $2\times4\times3$ cluster.
  }\label{FIG-NmtcN24}
\end{figure*}

\subsection{Scalar spin chirality}
Before disclosing patterns of dimer coverings, we wish to note that the underlying spin patterns support a magnetically ordered phase featuring finite scalar spin chirality.
On the one hand, if we define the magnetic order parameter as $M({\bf{Q}}) = \lim_{N\to\infty} \sqrt{\mathbb{S}_N({\bf{Q}})/N}$,
it is directly to find that \cite{Luo2022PRR}
\begin{align}
M^2(\boldsymbol{\Gamma'}) = \frac{S^2}{4} + \frac{\eta_a\eta_bab + \eta_b\eta_cbc + \eta_c\eta_aca}{2} = \frac{S^2}{4}
\end{align}
and
\begin{align}
M^2(\textbf{K}) = \frac{S^2}{6} - \frac{\eta_a\eta_bab + \eta_b\eta_cbc + \eta_c\eta_aca}{6} = \frac{S^2}{6},
\end{align}
where we have used the fact that $\eta_a\eta_bab + \eta_b\eta_cbc + \eta_c\eta_aca = 0$ since only one of the elements in $\{a, b, c\}$ is nonzero.
The above two equations also imply that $M(\textbf{K})/M(\boldsymbol{\Gamma'}) = \sqrt{6}/3 \approx 0.8165$,
which is verified by our DMRG calculation shown in Fig.~\ref{FIG-WpSSF}(b).
On the other hand, the scalar spin chirality is usually defined as \cite{Luo2022PRR,Yu2023arXiv}
\begin{equation}\label{EQ:ChiIJK}
{\chi}^{\triangle}_{ijk} = \left\langle\hat{\mathbf{S}}_i\cdot(\hat{\mathbf{S}}_j\times\hat{\mathbf{S}}_k)\right\rangle,
\end{equation}
where sites ($i, j, k$) form an equilateral triangle~($\triangle$) in the anticlockwise direction.
For the spins shown in the inset of Fig.~\ref{FIG-ChiralVBS}(a), it follows that
\begin{equation}\label{EQ:ChiIJK}
{\chi}^{\triangle}_{ijk} = (\eta_aa)^3 + (\eta_bb)^3 + (\eta_cc)^3 - 3\eta_a\eta_b\eta_c abc = S^3.
\end{equation}
The DMRG calculation of the scalar spin chirality presented in Fig.~\ref{FIG-ChiralVBS}(a) indeed suggests that it is close to 1 when away from the phase boundaries.
Thus, our results justify this spin texture as a magnetically ordered chiral spin state.

Finally, we turn to discuss the real space patterns of dimer coverings.
In the Cartesian states, there have and only have one sort of interaction for each bond,
favoring either Kitaev bond with an energy of $-K S^2$ or $\Gamma$ bond with an energy of $-\Gamma S^2$.
Due to the unique structures of the two interactions,
the $\Gamma$ bonds form a set of hexagonal loops while the Kitaev bonds connect these isolated hexagonal loops.
Therefore, this gives a columnar-like dimer pattern when Kitaev interaction is dominant and a plaquette-like dimer pattern otherwise,
see Fig.~\ref{FIG-ChiralVBS}(b) and Fig.~\ref{FIG-ChiralVBS}(c).
In the honeycomb lattice, we note that there is a dual (triangular) lattice which is constructed by placing a vertex inside each hexagon and connecting new vertices.
Thus, this forms a triangular Ising magnet with a three-sublattice $\{\eta_a, \eta_b, \eta_c\}$ unit cell.
For each equally-weighted hexagonal loop (which is centered at $\eta_b$ in Fig.~\ref{FIG-ChiralVBS}(b)
and $\eta_a$ in Fig.~\ref{FIG-ChiralVBS}(c)) in the two dimer patterns,
the nonzero component of each spin coincides with the type of outgoing bond,
rendering the corresponding plaquette operator $\langle\hat{W}_p\rangle$ be unitary.
By contrast, the plaquette operators at the remaining two cases are zero.
Therefore, the flux-like density $\langle \bar{W}_p \rangle = \langle \hat{W}_p \rangle/N_p$ ($N_p = N/2$) is found to be $1/3$.
Such a trimerization phenomenon is precisely confirmed by our DMRG calculation (see Fig.~\ref{FIG-WpSSF}(a)).
We note in passing that evidence of trimerization in spin-$1/2$ counterpart is also present,
albeit $\langle \overline{W}_p \rangle$ equals to $-1/3$ \cite{Rousochatzakis2020KITP}.
It is also worthwhile to note that these physical pictures should break down at the hidden SU(2) Heisenberg point where $K = \Gamma$ \cite{Chaloupka2015PRB}.

\section{Nematic Ferromagnets}\label{SEC:NmtcFM}

The aim of this Section is to unveil the nematic ferromagnets.
To this end, we start by using the von Neumann entropy, the hexagonal plaquette operator, and the Wilson loop operator to determine the QPTs.
Next, we pin down the nematic ferromagnets by examining nematic and magnetic order parameters.
Our findings are advocated by comparing results on 24-site hexagonal cluster and rhombic cluster.
Finally, we also present the landscape of bond energy on long cylinder at representative parameter points to verify the robustness of lattice nematicity.

\subsection{Nematic order parameter}
The FM Kitaev QSL is widely believed to be fragile against perturbations.
For example, the out-of-plane magnetic field of the strength $\sim 0.01 K$ can drive the ground state into the polarized phase \cite{Zhu2020PRR},
while the Heisenberg interaction of the strength $\sim 0.05 K$ can change the ground state into the long-range orderings \cite{Dong2020PRB}.
Therefore, one needs to resort sensitive quantities such as the entanglement entropy to probe the QPTs.
The entanglement entropy describes quantum correlations existed in the ground states and usually displays different behaviors when crossing quantum critical points.
Specifically, the von Neumann entropy is defined as $\mathcal{S}_{\rm vN}(l) = -\textrm{tr}\left(\rho_l\ln\rho_l\right)$
where $\rho_l$ is the reduced density matrix of the targeted subsystem with length $l$.
The length $l$ is typically fixed to be $N/2$ so that the blocks of the system and environment are equal in the DMRG calculation.

Figure~\ref{FIG-NmtcN24}(a) shows the von Neumann entropy $\mathcal{S}_{\rm vN}$ in the window of $\phi/\pi \in [0.92, 1.00]$ on the 24-site hexagonal cluster.
$\mathcal{S}_{\rm vN}$ has three discontinuities at $\phi_{3}^{\rm FM}/\pi \approx 0.952$, $\phi_{2}^{\rm FM}/\pi \approx 0.979$, and $\phi_{1}^{\rm FM}/\pi \approx 0.996$, respectively, signifying three first-order QPTs.
Noteworthily, the transition points are in accordance with these obtained from the flux-like density $\langle\overline{W}_p\rangle$.
While the emergent phases in the narrow region of $\phi_{3}^{\rm FM} < \phi < \phi_{1}^{\rm FM}$ are tempting,
it is challenging to disclose them for the lack of classical analogues.
Recalling that the SSFs shown in Fig.~\ref{FIG-WpSSF}(c) unambiguously indicate the broken of $C_3$ rotational symmetry,
it is thus reasonable to suspect that the lattice nematicity may arise.
To proceed further, we calculate the bond energy $E^{\gamma} = \langle H_{ij}^{\gamma}\rangle$ along the three Kitaev bonds.
As can be seen from the left panel in Fig.~\ref{FIG-NmtcN24}(b), there is an apparent difference in the bond energy in the region of $\phi_{3}^{\rm FM} < \phi < \phi_{1}^{\rm FM}$.
Typically, the absolute values of the bond energy along two of the three bonds are close and are bigger than the remaining one.
In our DMRG calculation along a specific path through the hexagonal cluster,
the weak bond is of the $\textbf{X}$-type when $\phi/\pi = 0.97$ and is of the $\textbf{Z}$-type when $\phi/\pi = 0.99$, see Fig.~\ref{FIG-NmtcN24}(c).
We thus define the nematic order parameter as the difference of the bond energy, i.e.,
$\mathcal{O}_{\rm NF} = \vert\min(E^{\gamma}) - \max(E^{\gamma})\vert$.
The $\mathcal{O}_{\rm NF}$ is indeed finite in the region of interest.
It seems to be smoothly evolve into the Kitaev QSL at $\phi_{1}^{\rm FM}$, while it undergoes a jump at $\phi_{3}^{\rm FM}$.

\begin{figure}[!ht]
\centering
  \includegraphics[width=0.95\columnwidth, clip]{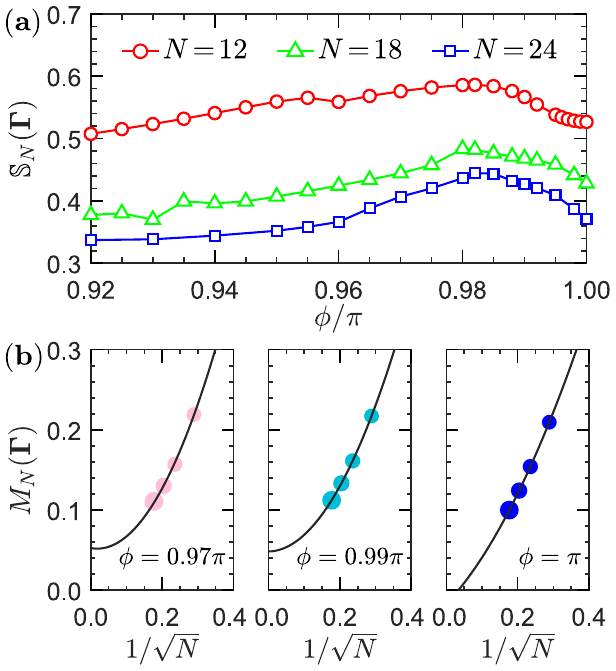}\\
  \caption{(a) The SSF $\mathbb{S}_N(\boldsymbol{\Gamma})$ as a function of $\phi/\pi$ on three $2\times L_x\times3$ clusters with $L_x$ = 2 ($N$ = 12, red circle), 3 ($N$ = 18, green triangle), and 4 ($N$ = 24, blue square).
  (b) Extrapolations of the FM order parameter $M_N(\boldsymbol{\Gamma})$ with respect to four clusters of $2\times2\times3$ ($N = 12$), $2\times3\times3$ ($N = 18$), $2\times4\times3$ ($N = 24$), and $2\times4\times4$ ($N = 32$). The parameter $\phi/\pi$ equals to 0.97 (left), 0.99 (middle), and 1.00 (right), and the error bars are less than the size of the symbols.
  }\label{FIG-NmtcFMSSF}
\end{figure}

\begin{figure*}[htb]
\centering
  \includegraphics[width=0.90\linewidth, clip]{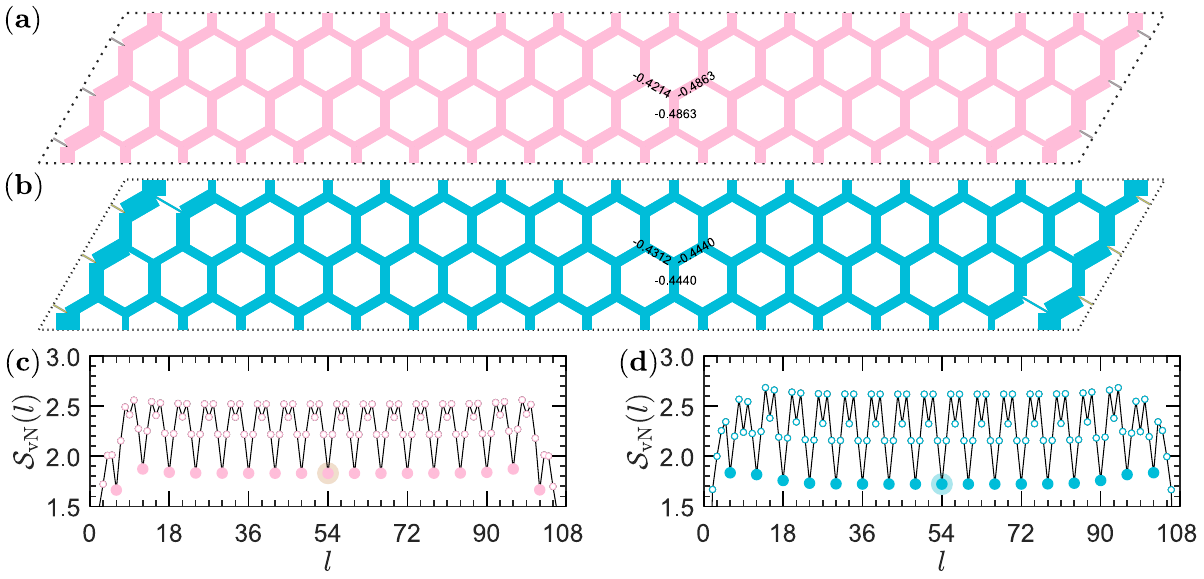}\\
  \caption{(a) and (b) show the landscapes of bond energy $E_{ij}$ on a $2\times18\times3$ long cylinder for $\phi/\pi$ = 0.97 and $\phi/\pi$ = 0.99, respectively.
  The thickness of the bond is proportional to cube of the relativity bond energy, i.e., $\propto (E_{ij}/\max(E_{ij}))^3$, for the guide of eyes.
  (c) and (d) show the von Neumann entropy $\mathcal{S}_{\rm vN}(l)$ of a consecutive segment of length $l$ on a $2\times18\times3$ long cylinder for $\phi/\pi$ = 0.97 and $\phi/\pi$ = 0.99, respectively.
  The solid symbols of the lowest branch represent the neat edge-cutting with $l$ being a multiply of six (i.e., the number of the sites along each column).
  }\label{FIG-NmtcCldBC}
\end{figure*}

To check for the possible cluster dependence, we compare the results to these obtained on a $2\times4\times3$ rhombic cluster, see Figs.~\ref{FIG-NmtcN24}(d)-(f).
Comparing Fig.~\ref{FIG-NmtcN24}(d) to Fig.~\ref{FIG-NmtcN24}(a), $\mathcal{S}_{\rm vN}$ also acquires three jumps or kinks
despite that the turning points are slightly different, i.e.,
$\tilde{\phi}_{3}^{\rm FM}/\pi \approx 0.962$, $\tilde{\phi}_{2}^{\rm FM}/\pi \approx 0.984$, and $\tilde{\phi}_{1}^{\rm FM}/\pi \approx 0.997$.
The Wilson loop operator $\langle W_l\rangle$ decreases gradually and goes down to zero when away from the Kitaev limit.
It has a steepest descent near $\tilde{\phi}_{1}^{\rm FM}$ and an evident jump at $\tilde{\phi}_{2}^{\rm FM}$,
followed by an inconspicuous hump with a maxima around $\tilde{\phi}_{3}^{\rm FM}$ (see the inset).
Pertaining the lattice nematicity, we also confirm that there is a difference in the three bond energy and one of the bond energy is obvious distinguished from the others.
However, it is the $\textbf{Y}$-type bond that always has the weakest strength in both nematic ferromagnets,
in contrast to the former case of $C_3$ hexagonal cluster.
Irrelevant to this difference, the maximum of $\mathcal{O}_{\rm NF}$ in the nematic ferromagnets are rather close,
with a value of 0.063 (hexagonal cluster) and 0.061 (rhombic cluster), respectively.
Nevertheless, the major difference lies in that the A$\Gamma$ phase also exhibits a finite lattice nematicity in the rhombic geometry,
as opposite to $C_3$-rotational hexagonal one.
This outcome is somewhat akin to its spin-$1/2$ analogy where the so-called nematic paramagnet is reported by the infinite DMRG calculation \cite{Gohlke2020PRR}.

\subsection{Magnetic order parameter}

A parallel question to address is the magnetic ordering in the nematic region.
According to the results on the 24-site hexagonal cluster,
the SSF has a primary peak at the $\boldsymbol{\Gamma}$ point and a competing peak at $\textbf{X}$ (or $\textbf{X}'$) point.
It is then curious to know how the SSF at $\boldsymbol{\Gamma}$ point will change as the system size varies.
Given the limitation of the DMRG calculation, we turn to utilize the rhombic clusters.
While these clusters are not equipped with the $C_3$ rotational symmetry of the Hamiltonian,
the $\boldsymbol{\Gamma}$ point with a zero momentum is always accessible, advocating the reliability of the magnetic ordering after a proper extrapolation.

Figure~\ref{FIG-NmtcFMSSF}(a) shows the $\phi$-dependence of SSF $\mathbb{S}_N(\boldsymbol{\Gamma})$
on three $2\times L_x\times3$ clusters with $L_x$ = 2 (red circle), 3 (green triangle), and 4 (blue square).
$\mathbb{S}_N(\boldsymbol{\Gamma})$ has a hump in the nematic regions and its value decreases consecutively with the increase of $N$ from 12 to 24.
In the magnetically ordered states, the magnetic order parameters usually behalve as $M \simeq c_0 + {c_1}/{\sqrt N}+{c_2}/{N}+ \cdots$
where $c_i$ ($i = 0, 1, 2$) are coefficients \cite{SchulzJPI1996}.
To proceed further, we make a quadratic polynomial extrapolation of the magnetic order parameter
$M_N(\boldsymbol{\Gamma}) = \sqrt{\mathbb{S}_N(\boldsymbol{\Gamma})/N}$ versus $1/\sqrt{N}$, see Fig.~\ref{FIG-NmtcFMSSF}(b).
We note that the magnetic order parameter on the $2\times4\times4$ cluster with $N = 32$ is taken into account, yet with the largest error bar.
The extrapolations at $\phi/\pi$ = 0.97 (left) and 0.99 (middle) in the nematic regions yield small but finite values around 0.05.
As a comparison, $M_N(\boldsymbol{\Gamma})$ at $\phi/\pi = 1$ in the QSL phase is estimated to be zero within the numerical precision.
Therefore, we conclude that the nematic regions indeed possess FM ordering and are thus termed nematic ferromagnets.

\begin{table}[th!]
\caption{\label{Tab-NmtcOP}
The nematic order parameter $\mathcal{O}_{\rm NF}$ in the two nematic ferromagnets at $\phi/\pi$ = 0.97 (NF-I) and 0.99 (NF-II).
The geometries are $C_3$-symmetric 24-site hexagonal cluster, $2\times4\times3$ rhombic cluster, and $2\times18\times3$ cylinder.}
\begin{ruledtabular}
\begin{tabular}{ c c c  c  c}
Phase   & Parameter             & $N = 24$      & $2\times4\times3$     & $2\times18\times3$    \\
\colrule
NF-I    & $\phi/\pi = 0.97$     & 0.0592        & 0.0615                & 0.0649                \\
NF-II   & $\phi/\pi = 0.99$     & 0.0158        & 0.0139                & 0.0128                \\
\end{tabular}
\end{ruledtabular}
\end{table}

\subsection{Nematicity on long cylinder}

Previously, we have unveiled a nematic region in the vicinity of the FM Kitaev QSL on two different 24-site clusters.
It is thus imperative to check for the tendency of nematicity on larger system size.
In this regard, we calculate the nearest-neighbor bond energy on a $2\times18\times3$ cylinder
where open boundary condition is employed on $\textbf{a}_1 (1, 0)$ direction, see Figs.~\ref{FIG-NmtcCldBC}(a) and (b).
The difference in the thickness of the three bonds clearly demonstrates a lattice nematicity at $\phi/\pi$ = 0.97 and 0.99,
and the intensity contrast is more conspicuous in the former than the latter.
When $\phi/\pi$ = 0.97, the DMRG calculation suggests that the bond energy in the middle region of the cylinder along the $\textbf{X}$, $\textbf{Y}$, and $\textbf{Z}$ bonds are $-0.4863$, $-0.4214$, and $-0.4863$, respectively, yielding a nematic order parameter of 0.0649.
As a comparison, the nematic order parameter at $\phi/\pi$ = 0.99 is only 0.0128, nearly one fifth of the former.
Despite the difference in intensity, the values of the nematic order parameters are exceedingly close to those obtained on 24-site clusters (cf. Fig.~\ref{FIG-NmtcN24}).
The values of the nematic order parameter $\mathcal{O}_{\rm NF}$ in the two nematic ferromagnets at representative points are shown in Table~\ref{Tab-NmtcOP}.
Despite of different cluster shapes and system sizes, the values are fairly close in each phase and should survive in the thermodynamic limit.
The robustness of the nematicity unambiguously emphasizes the lattice-rotational symmetry breaking in the nematic region.
In addition, geometry of the $2\times18\times3$ cylinder is commensurate to the 4-site zigzag order and 12-site order,
which peak at $\textbf{M}$ point and $\textbf{K/2}$ point in the first Brillouin zone, near the FM Kitaev limit at the classical level \cite{Rayyan2021PRB,Chen2023NJP}.
Our DMRG calculation suggests that the dominant peak is located at $\boldsymbol{\Gamma}$ point instead and thus does not favor these large-unit-cell ordering,
showing that the nematicity is a consequence of quantum effect.

To further reveal the nature of the nematic ferromagnets,
we present the behaviors of von Neumann entropy $\mathcal{S}_{\rm VN}(l)$ on a $2\times18\times3$ cylinder at $\phi/\pi$ = 0.97 and 0.99 in Figs.~\ref{FIG-NmtcCldBC}(c) and (d), respectively.
The curves are quite alike and the lowest branches in which $l$ is a multiply of six are fairly flat, indicative of a vanishing central charge $c$ in conformal field theory.
In other words, the excitations in the nematic ferromagnets seem to be gapped.
However, the difference of the two may lie in the distribution of entanglement spectrum.
It is found that the nematic ferromagnet I at $\phi/\pi$ = 0.97 is characterized by a nearly doubly degenerate entanglement spectrum at least for the lowest part,
while the entanglement spectrum in the nematic ferromagnet II at $\phi/\pi$ = 0.99 is odd (not shown).
Nevertheless, such a difference does not seem to be a universal character as it is not observed in the 24-site clusters.
Beyond the breaking of same symmetries, i.e., lattice-rotational symmetry and time-reversal symmetry,
our results imply that the intrinsic divide between the two nematic ferromagnets is still obscure and calls for future study.

\section{Conclusion}\label{SEC:CONC}

We studied the quantum phase diagram of the spin-1 Kitaev-$\Gamma$ model on the honeycomb lattice.
Through DMRG calculations on various clusters, we showed that the model harbours six distinct phases,
which are the FM and AFM Kitaev QSLs, the possible nonmagnetic A$\Gamma$ phase that embraces the pure $\Gamma$ limit,
the chiral spin state, and two nematic ferromagnets.
The Kitaev QSLs in the absence of $\Gamma$ term are characterized by a unitary hexagonal plaquette operator $\langle\hat{W}_p\rangle$ and Wilson loop operator $\langle\hat{W}_l\rangle$, and the $\Gamma$-driven QPTs can be captured by the singularities in $\langle\hat{W}_p\rangle$ and $\langle\hat{W}_l\rangle$.
In accordance with the spin-1 Kitaev-$\Gamma$ chain \cite{LuoPRR2021},
our result also indicates that the AFM Kitaev QSL is more robust against the $\Gamma$ interaction than its FM analogy.

Beyond these nonmagnetic phases, the main findings of this work are two categories of magnetically ordered states.
To begin with, when the Kitaev and $\Gamma$ interactions are both positive,
the chiral spin state is selected from the infinitely degenerate manifold of all possible states via order-by-disorder mechanism.
As a result, the ground state exhibits a noncoplanar pattern in which all spins point along cubic axes in the spin space.
It has a six-sublattice unit cell and displays peaks at $\textbf{K}$ and $\boldsymbol{\Gamma}'$ points in the Brillouin zone.
Noteworthily, this phase exhibits an almost saturated scalar spin chirality, signifying the broken of time-reversal symmetry.
As the Kitaev and $\Gamma$ interactions are different,
the real-space bond energy displays columnar-like (when $K > \Gamma$) or plaquette-like (when $K < \Gamma$) dimer pattern.

In addition, there are two emergent nematic ferromagnets in the vicinity of the dominating FM Kitaev regime.
These phases feature multiple-$Q$ structures where several Bragg peaks coexist in the reciprocal space.
Among these peaks, the one at $\boldsymbol{\Gamma}$ point seems to have the strongest intensity in a wide interval
and the proper extrapolations of this magnetic order parameter yield small but finite values,
indicative of magnetically ordered states.
In addition, the landscapes of the nearest-neighbor bond energy show apparent nonequivalence among the three bonds,
in which two of the bond energy are nearly equal but their absolute values are much larger than the remaining.
The resulting nematic order parameter defined as the difference in the bond energy keeps finite in a certain region
and the values are robust even on long cylinders.
We point out that the values are not satisfactorily large when compared to the bond energy,
and we speculate that they could be enhanced by other interactions or magnetic field, as studied in the spin-$1/2$ models \cite{Lee2020NC,Gohlke2020PRR}.
However, the jumps in von Neumann entropy and other quantities clearly demonstrate a first-order QPT in between.

While the differences in the preferred bond and entanglement spectrum are possible evidences to reveal the distinction,
decisive characters to distinguish the two remains unclear.
On all accounts, these results are unambiguous to identify two nematic ferromagnets as the ground states near the FM Kitaev QSL.
Therefore, our work may provide insights of the controversial region in the spin-$1/2$ model \cite{Rousochatzakis2024RoPP},
or underscore the peculiar property that is parasitic on the spin-1 system.
In the latter case, it is constructive to study the quadrupolar correlations to recognize the possible high-rank orderings.
For example, while the ground state of the spin-1 Kitaev chain is known to be nonmagnetic,
a recent work points out that it seems to be a spin-nematic phase by calculating the spin-nematic correlation \cite{LuoPRB2023}.
To conclude, we expect our study will stimulate further thorough investigations of novel phases in higher-spin models
and facilitate more synthesis of higher-spin Kitaev magnets.

\begin{acknowledgments}
We are indebted to H.-Y. Kee for the previous collaboration on related works.
This work is supported by the National Program on Key Research Project (Grant No. MOST2022YFA1402700),
the Natural Science Foundation of Jiangsu Province (Grant No. BK20220876),
the National Natural Science Foundation of China (Grants No. 12304176, No. 12274187, No. 12247183, and No. 12247101).
Q.L. also acknowledges the startup Fund of Nanjing University of Aeronautics and Astronautics (Grant No. YAH21129).
The computations are partially supported by High Performance Computing Platform of Nanjing University of Aeronautics and Astronautics.
\end{acknowledgments}



\appendix

\section{Static Structure Factor of the A$\Gamma$ phase}\label{appendixA}
In this Appendix, we showcase the static structure factor (SSF) of the A$\Gamma$ phase,
and give an argument for its possible nonmagnetic property mentioned in the main text.

\begin{figure}[!ht]
	\centering
	\includegraphics[width=0.98\columnwidth]{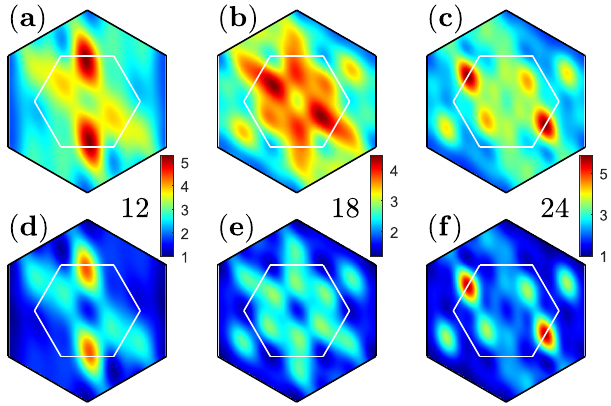}\\
	\caption{The scaled SSF $\tilde{\mathbb{S}}_N({\bf{q}})$ for the spin-$1/2$ [(a)-(c) in upper panels] and spin-1 [(d)-(f) in lower panels] A$\Gamma$ phase at $\phi/\pi = 0.5$.
    (a) and (c), (b) and (e), and (c) and (f) are for 12-site ($2\times2\times3$), 18-site ($2\times3\times3$), and 24-site ($2\times4\times3$) rhombic clusters, respectively.
    At each system size, the value of the scaled SSF in the spin-1 case is slightly smaller than that of the spin-$1/2$ (see the Appendix for numeric values).
	}\label{FIG-appendixA}
\end{figure}

The SSF is generally expected to capture the ordering wavevector for magnetically ordered states or displaying diffusive structure otherwise.
However, special caution should be paid when dealing with finite-size systems, since deceptive peaks may occur in the reciprocal space.
Before presenting the numerical results, we note that the classical ground state of the honeycomb $\Gamma$ model is a classical spin liquid \cite{Rousochatzakis2017PRL},
which has a highly degenerate manifold comprising the 4-site zigzag order (peaks at $\textbf{M}$ point) and 18-site triple-meron crystal (peaks at $2\textbf{M}/3$ point).
At the quantum level, we recall that there is a hot debate on the nature of A$\Gamma$ phase in the spin-$1/2$ honeycomb $\Gamma$ model
(for details, see Fig. 5 in Ref.~\cite{Rousochatzakis2024RoPP}).
Although proposals of zigzag order and incommensurate order (its peak is close to $2\textbf{M}/3$ point) are reported,
many other studies, including the large-scale DMRG \cite{Luo2021NPJ} and infinite DMRG calculations \cite{Gohlke2020PRR},
give a nonmagnetic phase albeit with different explanations.
To compare the spin-1 A$\Gamma$ phase to that of the spin-$1/2$, we introduce the scaled SSF $\tilde{\mathbb{S}}_N({\bf{q}})$,
which is given by $\mathbb{S}_N({\bf{q}})/S^2$ where $\mathbb{S}_N({\bf{q}})$ is the conventional SSF.

Figure~\ref{FIG-appendixA} shows the scaled SSF $\tilde{\mathbb{S}}_N({\bf{q}})$
for the spin-$1/2$ and spin-1 A$\Gamma$ phase at $\phi/\pi = 0.5$ in three upper and lower panels, respectively.
As the system size $N$ increases from 12, to 18 and to 24,
the scaled SSF has a soft peak at $\textbf{M}$ point (when $N$ is a multiple of 4) or $2\textbf{M}/3$ point (when $N$ is 18).
The fact that the leading peak varies for different system sizes originates from the competing states and manifests the fragility of magnetic ordering.
In addition, the values of the scaled SSF in the spin-1/2 case are 5.2760, 4.4636, and 5.8514,
while they are 4.4076, 2.9109, and 5.6222 in the spin-1 case.
Therefore, the scaled SSF, together with the scaled magnetic order parameter $\tilde{M}_N({\bf{Q}}) = \sqrt{\tilde{\mathbb{S}}_N({\bf{Q}})/N}$,
is even smaller than that of the spin-$1/2$ at each system size.
Similar calculations have also been performed in the A$\Gamma$ phase at $\phi/\pi = 0.6$, and the outcome remains the same.
As a result, although we admit that the SSF at small system sizes cannot draw a conclusion with certainty,
the fact that the scaled SSF undergoes a similar behavior to that of the spin-$1/2$ case
may indicate that the ground state of the spin-1 A$\Gamma$ phase is also likely a nonmagnetic phase.





%

\end{CJK*}
\end{document}